\documentclass{article}

\usepackage{PRIMEarxiv}

\usepackage[utf8]{inputenc} 
\usepackage[T1]{fontenc}    
\usepackage{float}
\usepackage{hyperref}       
\usepackage{url}            
\usepackage{booktabs}       
\usepackage{amsfonts}       
\usepackage{nicefrac}       
\usepackage{microtype}      
\usepackage{lipsum}
\usepackage{fancyhdr}       
\usepackage{graphicx}       
\usepackage{amsmath}
\graphicspath{{media/}}     
\usepackage[table]{xcolor}
\usepackage{adjustbox}  
\usepackage[caption=false,font=normalsize,labelfont=sf,textfont=sf]{subfig}
\usepackage[square,sort,comma,numbers]{natbib}
\usepackage{algorithm,algorithmic}

\usepackage[scaled]{helvet} 
 
\usepackage{mathpazo} 
\title{CSIM: A Copula-based similarity index sensitive to local changes for Image quality assessment}
\author{
  \href{https://orcid.org/0000-0002-5403-3911}{\includegraphics[scale=0.1]{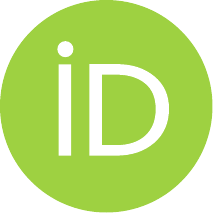}\hspace{1mm} Safouane EL GHAZOUALI*}\\
  TOELT LLC AI lab\\
  Winterthur, Swintzerland \\
  \texttt{safouane.elghazouali@toelt.ai} \\
  \And
  \href{https://orcid.org/0000-0002-6060-5365}{\includegraphics[scale=0.1]{orcid.pdf}\hspace{1mm}Umberto MICHELUCCI} \\
  TOELT LLC AI lab, \\
  Winterthur, Swintzerland \\
  \texttt{umberto.michelucci@toelt.ai} \\
  \And
  \href{https://orcid.org/0000-0002-3980-9902}{\includegraphics[scale=0.1]{orcid.pdf}\hspace{1mm}Yassin EL HILLALI}\\
  Polytechnique Haut-de-France, \\
  Valenciennes, France \\
  \texttt{yassin.elhillali@uphf.fr}\\
  \And
  \href{https://orcid.org/0000-0001-8696-342X}{\includegraphics[scale=0.1]{orcid.pdf}\hspace{1mm}Hichem NOUIRA}\\
  LNE-CNAM, \\
  Paris, France \\
  \texttt{hichem.nouira@lne.fr}\\
}

\begin{document}
\maketitle

\begin{abstract}
Image similarity metrics play an important role in computer vision applications, as they are used in image processing, computer vision and machine learning. Furthermore, those metrics enable tasks such as image retrieval, object recognition and quality assessment, essential in fields like healthcare, astronomy and surveillance. Existing metrics, such as PSNR, MSE, SSIM, ISSM and FSIM, often face limitations in terms of either speed, complexity or sensitivity to small changes in images. To address these challenges, a novel image similarity metric, namely CSIM, that combines real-time while being sensitive to subtle image variations is investigated in this paper. The novel metric uses Gaussian Copula from probability theory to transform an image into vectors of pixel distribution associated to local image patches. These vectors contain, in addition to intensities and pixel positions, information on the dependencies between pixel values, capturing the structural relationships within the image. By leveraging the properties of Copulas, CSIM effectively models the joint distribution of pixel intensities, enabling a more nuanced comparison of image patches making it more sensitive to local changes compared to other metrics. Experimental results demonstrate that CSIM outperforms existing similarity metrics in various image distortion scenarios, including noise, compression artifacts and blur. The metric's ability to detect subtle differences makes it suitable for applications requiring high precision, such as medical imaging, where the detection of minor anomalies can be of a high importance. The results obtained in this work can be reproduced from this Github repository: \url{https://github.com/safouaneelg/copulasimilarity}.
\end{abstract}

\newpage

\section{Introduction}
Similarity metrics in computer vision play an important role in assessing the effectiveness and performance of various image and video processing algorithms \citep{9358789}. These metrics serve as quantitative measures that allow the comparison of the output for different algorithms against ground truth data or among newly proposed methodologies. The ability to quantify the similarity between two images or videos is useful in a wide range of industries (healthcare \citep{app12083754}, metrology \citep{10.1117/12.2614281}, autonomous driving \citep{wang2023evaluationoptimizationrenderingtechniques}).

Over the years, image similarity metrics have been proposed and compared, demonstrating strengths and weaknesses. The most popular metrics are Structural similarity index (refered to as SSIM) \citep{1284395}, Mean Squared Error (MSE) and Peak Signal to Noise Ratio (PSNR) \citep{Kudlka2012ImageQA}. The selection of an appropriate similarity metric remains a challenging task, as different metrics may yield varying results depending on the specific application and usage. For example, SSIM provide a score varying from 0 (no similarity found) to 1 (identical images), while PSNR gives a score in decibels (dB) that represents the ratio of the maximum possible power of a signal to the power of corrupting noise. Therefore, a thorough understanding of the underlying principles and characteristics of each similarity metric is essential in selecting the most suitable metric for a particular task \citep{8342690, 8369657}.

The development of objective image quality assessment metrics has focused on predicting perceived image structure, features and form. These metrics can monitor and adjust image quality, optimize image processing systems and benchmark deep learning models performance \citep{moser2024diffusionmodelsimagesuperresolution, reddi2019mlperf}. Existing metrics often overlook the complex dependencies between pixels, relying on structural data or features such as edge and corners. This limitation can lead to inaccurate similarity measurements, particularly when subtle changes and distortions between images can be observed.

In this paper, one new approach based on Gaussian Copula formula \citep{Nelsen2006IntroductionToCopulas} is proposed to measure image similarity. Copula-based SIMilarity index (namely CSIM) allows the analysis of relationships between pixel intensities in small local patches. This method is particularly useful for detecting minor and localized changes or distortions between images, which is considered a valuable tool for applications like image registration/stitching and quality assessment. Furthermore, this metric can also be used as a more robust loss function for deep learning model such as synthetic image generation \citep{8297089}, diffusion models \citep{10287612} and deepfake detection \citep{9964852}. The advantage of Gaussian copulas remains on their capacity of capturing complex dependencies between pixels and thus a more accurate prediction of similarity between two images, rather than relying on structural data or features such as shapes, edges and corners.

The contributions of this paper are presented as following:
\begin{itemize}
    \item Mathematical model for sensitive similarity metric based on Gaussian Copula.
    \item Experimental comparisons of the proposed metric to existing ones including SSIM, ISSM and FSIM on static cases (quality assessment on two images) and dynamic one (image to image comparison in video with a large unchanged background).
    \item Large scale metric assessment on a dataset that consists of 30 original images and a total of 900 distorted versions.
    \item Complexity analysis on the implemented CSIM algorithm while testing a range of value for the patch size hyper-parameter.
\end{itemize}

Therefore, this paper is structured as follow: section \ref{sec:sota} presents an overview of the existing metrics for comparing two images. Section \ref{sec:copula-theory} presents the theoretical backbone of the proposed CSIM approach. An experimental study is given in section \ref{sec:Experimentalstudy} where the copula similarity is applied on real and synthetically distorted images and than compared to other existing metrics. Finally, the results are presented and discussed in section \ref{sec:result_discussion} along with an analysis of algorithmic complexity of the proposed CSIM.

\section{Similarity metrics: state-of-the-art}\label{sec:sota}

\subsection{Structural Similarity Index (SSIM)}

SSIM \citep{nilsson2020understandingssim, BAKUROV2022116087, 1284395} is the most widely used metric for evaluating the structural similarity between two images. It was first introduced by Wang \textit{et. al.} \citep{1284395} as an efficient alternative to the traditional metrics such as MSE \citep{sara2019image} and PSNR \citep{5596999}. SSIM judges the visual similarity between two given images by comparing their parameters of luminance, contrast and structural features. This metric has been extensively used in various applications, including image and video compression \citep{BAKUROV2022116087, cheng2019learningimagevideocompression}, image denoising \citep{nilsson2020understandingssim}, image quality assessment \citep{1284395, 5596999} and image registration \citep{Amintoosi2009}. Additionally, SSIM has been employed in the evaluation of image processing algorithms, such as image perceptual down scaling \citep{10.1145/2766891} and image restoration \citep{7797130}. The effectiveness of SSIM in capturing the human visual perception of image similarity has made it also one popular choice as loss function for training deep learning models such as image restoration \citep{PEKMEZCI2024104685, 7797130} and realistic image generation \citep{czolbe2021lossfunctiongenerativeneural}.

Given two image to compare, the SSIM metric is calculated using the formula expressed in Eq. \ref{eq:ssim}:

\begin{equation}\label{eq:ssim}
SSIM({I_1}, {I_2}) = \frac{(2\mu_{I_1}\mu_{I_2} + C_1)(2\sigma_{{I_1}{I_2}} + C_2)}{(\mu_{I_1}^2 + \mu_{I_2}^2 + C_1)(\sigma_{I_1}^2 + \sigma_{I_2}^2 + C_2)}
\end{equation}

\noindent where:
\begin{itemize}
    \item ${I_1}$ and ${I_2}$ are the two input images
    \item $\mu_{I_1}$ and $\mu_{I_2}$ are their respective means
    \item $\sigma_{I_1}$ and $\sigma_{I_2}$ are their respective standard deviations
    \item $\sigma_{{I_1}{I_2}}$ is the covariance between ${I_1}$ and ${I_2}$
    \item $C_1$ and $C_2$ are constants used to stabilize the division
\end{itemize}

One advantage of using SSIM over MSE (or RMSE) and PSNR is its ability to capture structural information of an image \citep{nilsson2020understandingssim}, which is closely related to human visual perception. Additionally, SSIM is more robust to variations in image intensity and contrast in comparison to MSE and PSNR, making it a reliable metric for image quality assessment. Nevertheless, even though SSIM can locate intensity differences between two images on a large scale, it fails to compare localized and small changes within images with a broad and uniform background, making it one of its well-known limitations \citep{7351345}. Furthermore, according to Dohmen \textit{et. al.} study \citep{dohmen2024pitfallsassessingsyntheticmedical} on similarity metrics pitfalls in the medical imagery, SSIM may not be suitable for images with complex textures or patterns, since it also returns a high similarity scores with misaligned images. The SSIM also fails to identify blurring pattern at the pixel-level as more blurring leads to higher similarity score.

Despite these limitations, SSIM has inspired the development of several derivatives, including Multi-Scale Structural SIMilarity (MS-SSIM) \citep{1292216}, which incorporates image details at different resolutions by iteratively applying a low-pass filter and downsampling the filtered image and combining the measurements at different scales using a weighted sum. Another derivative of classical SSIM is the Complex Wavelet Structural SIMilarity (CW-SSIM). It is an extension of the SSIM method to the complex wavelet domain designed to be insensitive to non-structural geometric image distortions, such as translation, scaling and rotation, while being sensitive to structural distortions (JPEG compression). The CW-SSIM index separates the measurement of magnitude and phase distortions and is more sensitive to phase than magnitude distortions.

\subsection{Feature Similarity Index (FSIM)}
L. Zhang \textit{et. al.} introduced FSIM \citep{5705575}, a metric based on the similarity of low-level features between a reference image and a distorted image. By incorporating chrominance information index, FSIM can also capture differences in colorful images. FSIM is designed to measure the similarity between images based on phase congruency (PC) \citep{7295942, WANG20112015} and gradient magnitude (GM) \citep{ADAMO20091304} features.

The FSIM index is computed in two stages: (1) the local similarity map is computed by comparing the PC and GM features between the reference and distorted images. The PC feature is used to capture the structural information in the image, while the GM feature is used to capture the contrast information. (2) The local similarity map is then pooled into a single similarity score using a weighting function based on the PC feature.

The evaluation of the FSIM index has been carried out on benchmark databases and compared with state-of-the-art methods. The experimental results show that FSIM outperform the other methods in terms of consistency with human subjective evaluations \citep{5705575}. The FSIM method work as the following: (1) The PC feature (Eq. \ref{eq:phasecongr}) is computed using a log-Gabor filter, which is a biologically plausible model of how mammalian visual systems detect and identify features in an image. (2) the GM (Eq. \ref{eq:gradmagni}) feature is computed using a gradient operator, such as the Sobel operator.

\begin{equation}\label{eq:phasecongr}
    PC(x) = \frac{E(x)}{A(x) + \epsilon}
\end{equation}

\noindent where $E(x)$ is the local energy, $A(x)$ is the local amplitude and $\epsilon$ is a small positive constant to avoid zero division.

\begin{equation}\label{eq:gradmagni}
    GM(x) = \sqrt{G_x(x)^2 + G_y(x)^2}
\end{equation}

\noindent $G_x(x)$ and $G_y(x)$ are the partial derivatives of the image along the horizontal and vertical directions, respectively.

The FSIM index is computed by combining the PC and GM features using a similarity measure. The similarity measure is defined by Eq. \ref{eq:similmeasure}.

\begin{equation}\label{eq:similmeasure}
    S(x) = \frac{2PC_1(x)PC_2(x) + T_1}{PC_1(x)^2 + PC_2(x)^2 + T_1} \cdot \frac{2GM_1(x)GM_2(x) + T_2}{GM_1(x)^2 + GM_2(x)^2 + T_2}
\end{equation}

\noindent $PC_1(x)$ and $PC_2(x)$ are the PC features of the reference and distorted images, respectively and $GM_1(x)$ and $GM_2(x)$ are the GM features of the reference and distorted images, respectively. $T_1$ and $T_2$ are positive constants.

The FSIM index is finally computed by pooling the local similarity map using a weighting function based on the PC feature (Eq. \ref{eq:fsim_eq}).

\begin{equation}\label{eq:fsim_eq}
    FSIM = \frac{\sum_{x \in \Omega} S(x) \cdot PC_m(x)}{\sum_{x \in \Omega} PC_m(x)}
\end{equation}

\noindent $PC_m(x) = \max(PC_1(x), PC_2(x))$ is the maximum PC value at location $x$ and $\Omega$ is the whole image spatial domain.

According to \citep{5705575}, the FSIM is effective in measuring image quality and outperform other state-of-the-art  methods in terms of consistency with human subjective evaluations.

\subsection{Information theoretic-based Statistic Similarity (ISSM)}

M. Aljanabi \textit{et. al.} proposed ISSM \citep{22797254.2019.1628617} aiming to reduce the challenges faced by image similarity measures, particularly in cases where images have low resolution, distortion, different lighting and background changes. The measure combines information theory and statistical approaches to provide a more accurate and reliable similarity metric. The foundation of ISSM lies in Shannon entropy theory \citep{cover1999elements} and joint histogram \citep{Pass1999}, which are used to construct a lossless image similarity-recognition measure.

The ISSM measure $I(x,y)$ between two images $x$ and $y$ is defined by Eq. \ref{eq:ISSM_sim}.

\begin{equation}\label{eq:ISSM_sim} 
    I(x,y)=\frac{C \cdot EHS \cdot (a+b)+e}{a \cdot C \cdot EHS+b \cdot EHS+c \cdot S(x,y)+e} 
\end{equation}

\noindent where $EHS$ is the Entropy-Histogram Similarity, $S$ represents SSIM and $C$ is the 2D edge correlation coefficient function of both image $x$ and $y$. The constants $a$, $b$ and $c$ are chosen to balance the quotient and avoid division by zero.

The Entropy-Histogram Similarity (EHS) is calculated between two images using Shannon entropy (Eq. \ref{eq:EHS_entropy}). 

\begin{equation}\label{eq:EHS_entropy} 
    EHS=- \sum_{k=1}^{M \times N} T(k) \log_2[T(k)] 
\end{equation}

\noindent where $T$ is the 2D joint histogram reshaped into a one-dimensional column vector.

The ISSM has been implemented and compared to SSIM and FSIM on image databases and delivered, under gaussian blurring and noising, better estimation of the similarity between images as it was more sensitive to local and global changes.

\section{Proposed method: Copula-based Similarity (CSIM)}\label{sec:copula-theory}

In probabilistic theory, copulas \citep{joe2014dependence} are used to understand and model complex dependencies between random variables. Unlike traditional multivariate distributions that are often limited by their parametric forms \citep{Sarabia2008, TACQ2010332, HARRAR2022104855}, copulas allow for the construction of multivariate distributions that can capture a wide range of dependency structures, making them highly adaptable to various applications \citep{SAHIN2024e28270, Yan2023}. The foundation of copula theory lies in Sklar's theorem \citep{math12030380}, which states that any multivariate joint distribution can be decomposed into its marginals and a copula while describing the dependence structure between the variables \citep{Nelsen2006IntroductionToCopulas}. This decomposition is particularly valuable in fields where understanding the joint behavior of variables is needed. One of the most well known applications of copulas is in finance \citep{cherubini2011dynamic}, such as risk management \citep{10.1002/for.3009}, portfolio optimization \citep{SAHAMKHADAM20221055, Sahamkhadam2021} and credit risk \citep{bouye2000copulas}.
\\
In the realm of image processing and computer vision, copulas has been employed in few applications related to areas such as image texture retrieval and analysis \citep{ETEMAD2021104256, 6777560}, pattern recognition \citep{TAMBORRINO2023100070} and change detection \citep{4481231}. By modeling the dependencies between pixel intensities or color channels, copulas can capture the fine spatial relationships that define textures and patterns in images. For example, in color texture retrieval, copulas are used to model the dependencies between different color channels and spatial scales, enabling the development of robust image retrieval algorithms that can effectively differentiate between textures \citep{LI2017118}. Therefore, copulas provide a means to understand and quantify the statistical dependencies that are not apparent within the image structure when considering each pixel or color channel independently. The theoretical background is details in the following section.

\begin{figure*}[ht]
    \centering
    \includegraphics[width=\textwidth, scale=1]{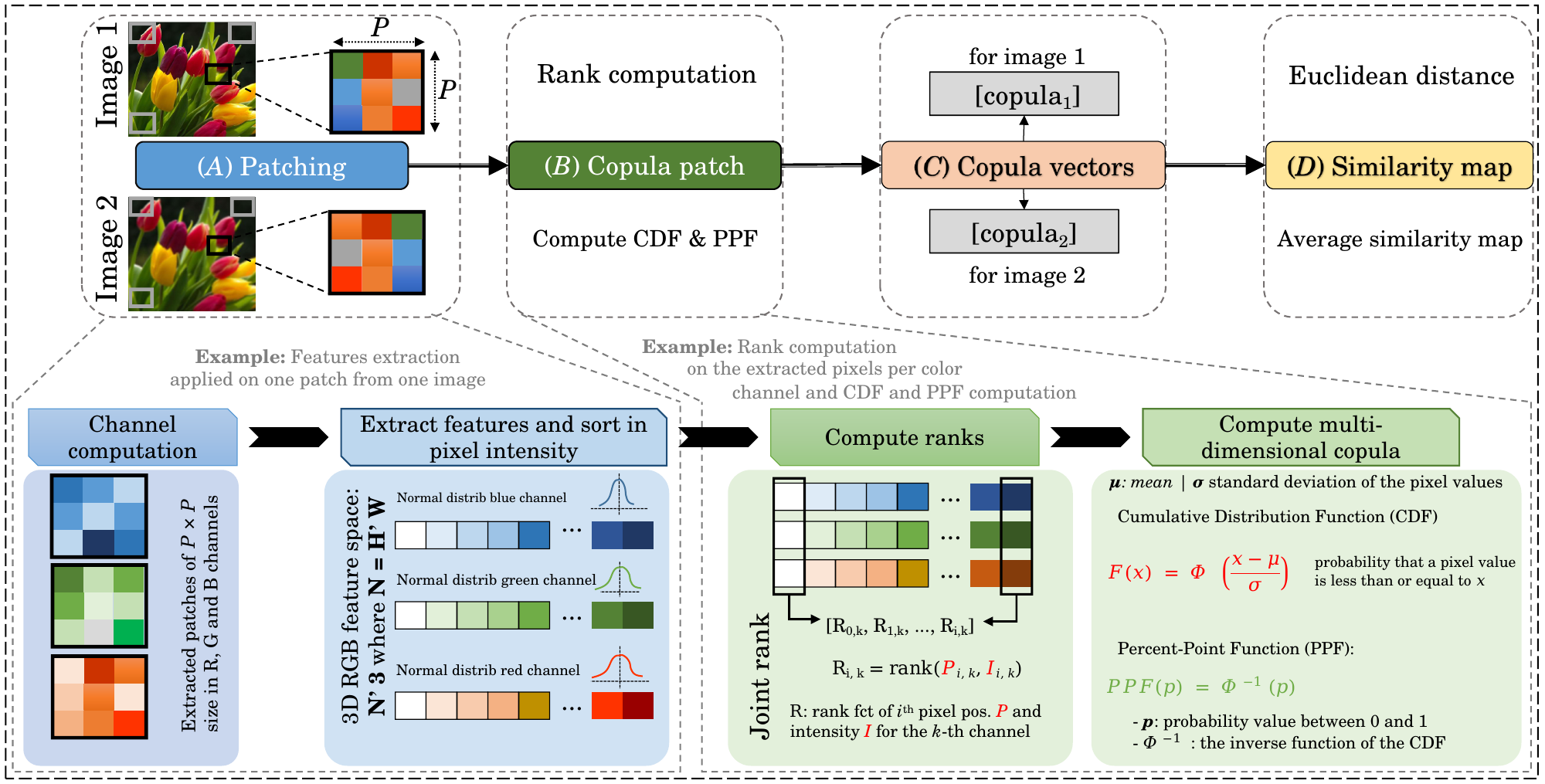}
    \caption{Flowchart of the copula-based similarity metric (CSIM) for image comparison distributed on four steps: (A) Patching: includes color channel computing, features extracting and sorting. (B) Copula patch: incorporates rank calculation, CDF and PPF. (C) Copula vectors: merge the copula patches into a vector for each image. (D) Similarity Map: Includes the calculated distances between the copula vectors and similarity scoring.}
    \label{fig:copula_similarity_chart}
\end{figure*}

In the context of image similarity computation, the proposed CSIM is used to compare images by analyzing local patches. The key idea is to evaluate how similar the statistical dependencies of pixel intensities are between corresponding patches in the two images. This allows for a detailed assessment of image similarity at a local scale, which makes the CSIM sensitive to localized changes and distortions (Flowchart summarized in 
 Fig. \ref{fig:copula_similarity_chart}).

The steps presented in the flowchart can be grouped into four main actions:

\begin{itemize}
    \item Patching and features sorting: 
    extract patches from each image, sort pixel intensities and create feature vectors to facilitate comparison.
    
    \item Compute Copula per image: 
    calculate the ranks and normalize them to capture the dependency structure within patches using copulas.
    
    \item Copula vectors: 
    combine copulas from all patches into a single vector representing the overall image characteristics.
    
    \item Euclidean distance: 
    measure similarity between two images by computing the Euclidean distance between their copula vectors and conversion into a similarity score.
\end{itemize}

\subsection{Patching and features sorting}

The first step in the process involves extracting non overlapping patches of a fixed size \( P \times P \) from each image. The patching of the images enables focusing the analysis on localized regions of interest within the images. Each image is sliced into a grid of patches, while ensuring that both images have the same resolution and patch size. The red, green and blue (RGB) color channels within the grid of each partition are represented by pixel values and transformed into features vectors by flattening them. The pixel intensities are then sorted in ascending order for subsequent statistical analysis. The sorting of the pixel intensities goal is to calculate the features ranks and the modeling of dependency structures through copulas, as it aligns the data to reveal underlying patterns and relationships.

\subsection{Image Copula computing}

Once the pixel intensities are sorted, the next step involves computing the copula for each image patch. This is performed by calculating the ranks for each pixel intensity vector within a patch. The rank  $R_{i,k}$ of the $k^{th}$ channel of pixel $i$ is determined by the order of pixel intensity values within the patch, considering both pixel position and intensity. The ranks are then normalized to the range [0,1] using Eq. \ref{eq:normalized_ranks}.

\begin{equation}\label{eq:normalized_ranks}
    \hat{R}_{i,k} = \frac{R_{i,k}}{N}
\end{equation}

\noindent $N$ represents the total number of pixels in the patch. Normalizing the ranks ensures that they are comparable across different patches and images. Afterwards, the cumulative distribution function (CDF) of the normalized ranks $\hat{R}$ is computed, assuming a Gaussian distribution of pixel intensities for each color channel. The CDF function \citep{6204462}, denoted as $\text{CDF}(\hat{R})$, provides a statistical measure on the probability of a pixel intensity (Eq. \ref{eq:CDF}).

\begin{equation}\label{eq:CDF}
    \text{CDF}(\hat{R}) = \int_{-\infty}^x f_{\hat{R}}(t) \, dt
\end{equation}

\noindent where $x$ refers to the normalized rank of the pixel intensities within a patch, $f_{\hat{R}}$ is the probability density function (PDF) of $\hat{R}$. After obtaining the CDF, the percent point function (PPF) (also known as Quantile function) \citep{Redivo2023} is applied to derive the copula, which captures the dependency structure between pixel intensities within a patch. The copula $C_{\hat{R}}(u)$ is expressed by the Eq. \ref{eq:copula_density}:

\begin{equation}\label{eq:copula_density}
    C_{\hat{R}}(u) = \text{PPF}(F_{\hat{R}}(u))
\end{equation}

\noindent $\text{PPF}(\cdot)$ being the inverse of the CDF for the normal distribution. This transformation maps the CDF to a standard normal distribution, allowing for a uniform comparison of dependencies across patches.

\subsection{Copula Vectors Generation}

The computed copulas for all patches are merged into a single vector representing the entire image. This aggregation is performed to facilitate a comprehensive analysis of the image based on the dependencies observed in each patch. The copula vector for an image is defined as $\mathbf{C}_\text{image} = [\mathbf{C}_{\text{patch}_1}, \mathbf{C}_{\text{patch}_2}, \ldots, \mathbf{C}_{\text{patch}_n}]$. \( \mathbf{C}_{\text{patch}_i} \) represents the copula for patch \( i \). By combining the copulas into a vector, it becomes possible to compare images based on their overall dependency structures, rather than individual pixel intensities or isolated image features.

\subsection{Euclidean distance calculation}

The final step in the process involves computing the Euclidean distance \citep{10.1137/120875909, 10.1007/978-3-540-30503-3_28} between the copula vectors of corresponding patches from the two images. The Euclidean distance provides a quantitative measure of dissimilarity between the copula vectors, reflecting the differences in dependency structures captured by the copulas. The Euclidean distance \( d(\mathbf{C}_1, \mathbf{C}_2) \) between two copula vectors \( \mathbf{C}_1 \) and \( \mathbf{C}_2 \) is calculated as:

\begin{equation}
    d(\mathbf{C}_1, \mathbf{C}_2) = \sqrt{\sum_{i=1}^{n} \left( \mathbf{C}_{1,i} - \mathbf{C}_{2,i} \right)^2}
\end{equation}

\noindent \( \mathbf{C}_{1,i} \) and \( \mathbf{C}_{2,i} \) are the copula vectors of the \( i \)-th patch from images 1 and 2, respectively. To convert the distance into a similarity score, the following formula is used (Eq. \ref{eq:euclidian-to-score}):

\begin{equation}\label{eq:euclidian-to-score}
    S = \max(0, 1 - \frac{d(\mathbf{C}_1, \mathbf{C}_2)}{\sqrt{n}})
\end{equation}

\noindent This formula ensures that the similarity score $S$ is within the range [0,1], where a score of 1 indicates high similarity (i.e., identical dependency structures) and a score of 0 indicates no similarity. The normalization by \( \sqrt{n} \) accounts for the dimensionality of the copula vectors, ensuring that the similarity score is not biased by the number of patches used in the comparison.

\section{Experiments}\label{sec:Experimentalstudy}

Several experimental studies have been conducted to assess the efficiency of the proposed CSIM metric. A comparative study is conducted between CSIM and existing similarity metrics including the FSIM, SSIM and the ISSM. Different types of images have been considered as well as artificially distorting images and comparing to the original ones.

\subsection{Performance evaluation under blurring and Gaussian noise}

\begin{figure}[!t]
    \centering
    \includegraphics[width=\textwidth]{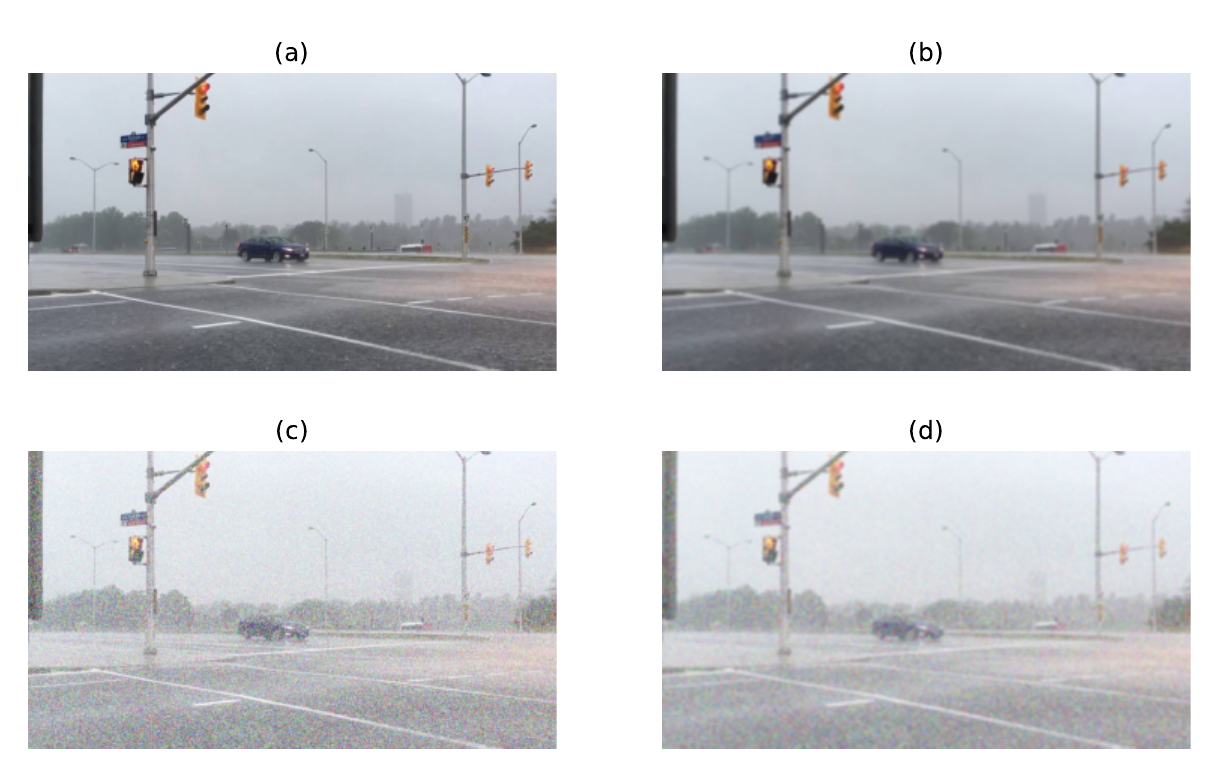}
    \caption{Visualization example of noise and blur application on an RGB image: (a) Original image, (b) addition of white Gaussian noise with a mean of 5 and standard deviation of 10, (c) blurred version of the noisy image with a standard deviation of 10 and kernel size of (5,5).}
    \label{fig:visualization_noise_blur}
\end{figure}

The performance of image quality metrics under several levels of blurring \citep{zhang2018learning} and Gaussian noise \citep{Ding_2021}, which are common forms of distortion in real-world scenarios, is conducted. To assess the impact of blurring, images were subjected to Gaussian blur with varying sigma values. Similarly, Gaussian noise with different standard deviation values was also applied to the images. The noise parameters were set as follows: noise means 5 and the standard deviations were varied between 0 and 20. Fig. \ref{fig:visualization_noise_blur} show an example of the original image and subsequent distorted versions. For each combination of noise and blur, the FSIM, SSIM, CSIM and ISSM metrics were computed.

\begin{figure*}[!t]
    \centering
    \subfloat[]{\includegraphics[width=0.9\textwidth]{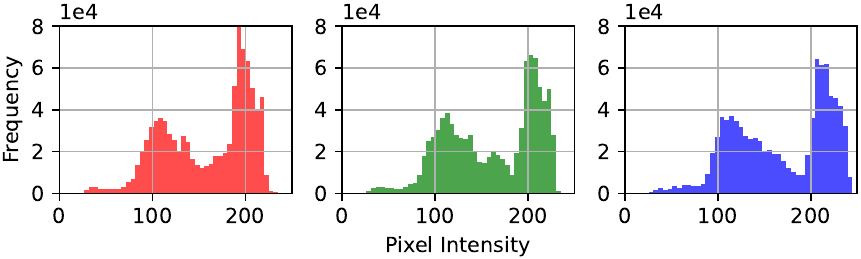}%
    \label{fig:pixel_intensity_distributions}}
    \hfil
    \subfloat[]
    {\includegraphics[width=0.9\textwidth]{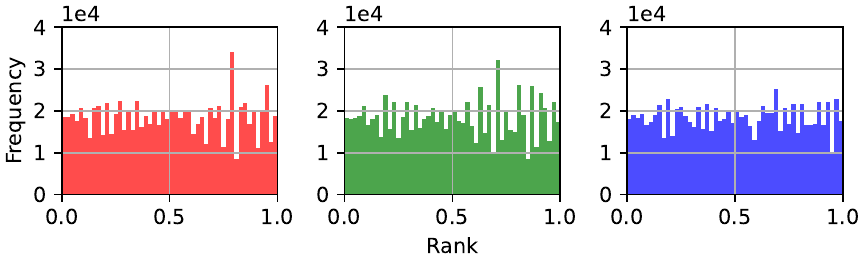}%
    \label{fig:ranks_distribution}}
    
    \subfloat[]
    {\includegraphics[width=0.9\textwidth]{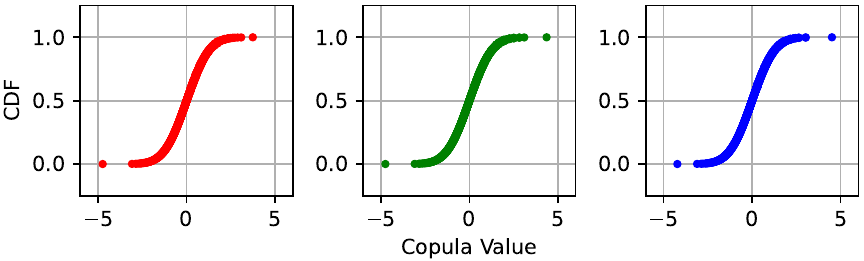}%
    \label{fig:cdf_distribution}}
    \hfil
    \subfloat[]{\includegraphics[width=0.9\textwidth]{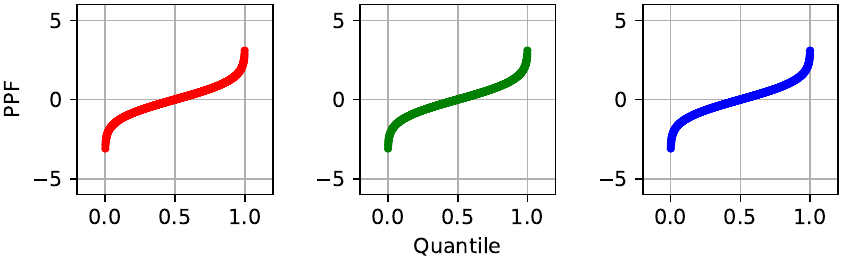}%
    \label{fig:ppf_distribution}}
    
    \caption{Visualization of feature extraction and ranking along with and copula vectors computation through CDF and PPF: (a) Histograms of the pixel intensity distributions for each color channel of the image. (b) Histograms of the ranks for each color channel. Statistical analysis of the copula-transformed pixel intensities for each color channel: (c) CDF of the red, green and blue channel copula. (d) PPF of the red, green and blue channel copula.}
    \label{fig:fig_sim}
\end{figure*}

To visualize the Copula transformation effect on the image, the statistical distributions of pixel intensities and the copula transformations for each color channel of the \textit{original image} have been analysed. Fig. \ref{fig:pixel_intensity_distributions} and \ref{fig:ranks_distribution} presents respectively histograms of the flattened pixel intensities and their respective ranks. Furthermore, Fig. \ref{fig:cdf_distribution} and \ref{fig:ppf_distribution} provides a statistical analysis by showcasing respectively the cumulative distribution functions (CDF) and the percent point functions (PPF) of the transformed pixel intensities. The CDF figures illustrate the distribution of copula values across different quantiles, while the PPF demonstrate the inverse CDF, providing a different perspective on the data distribution.

\begin{figure}[!t]
    \centering
    \subfloat[]{\includegraphics[width=0.7\textwidth]{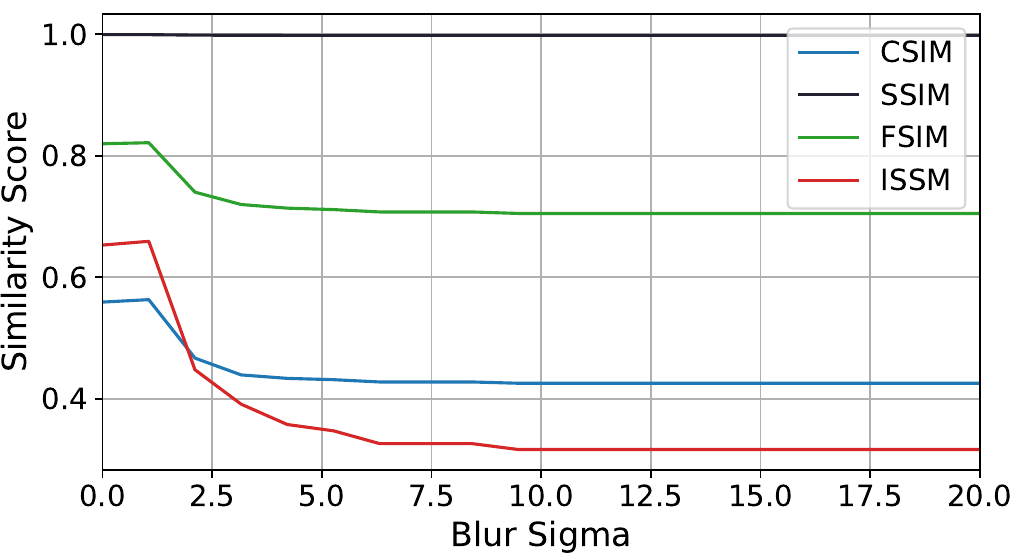}%
    \label{fig:metrics_vs_blur}}
    \hfil
    \subfloat[]{\includegraphics[width=0.7\textwidth]{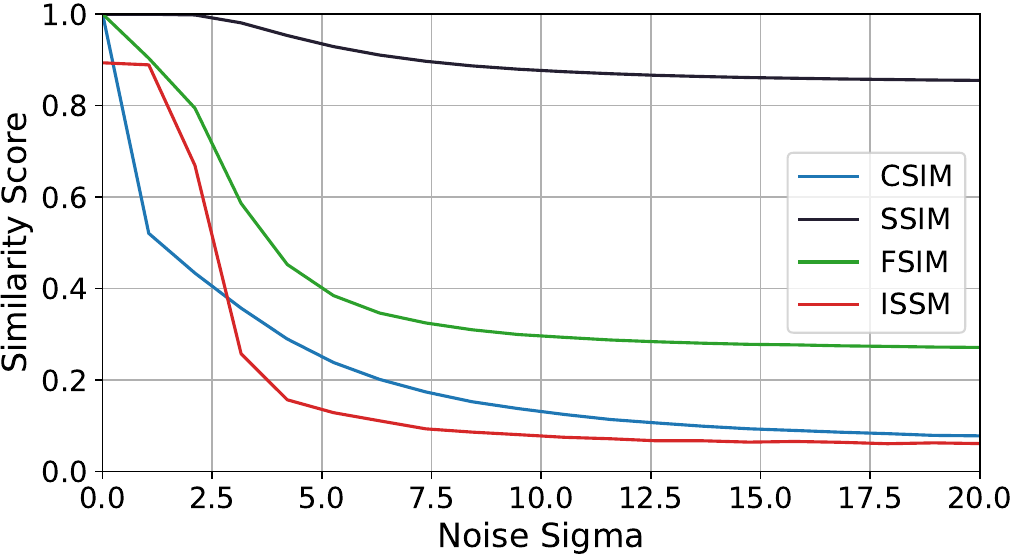}%
    \label{fig:metrics_vs_noise}}
    \caption{Comparison of image quality metrics under varying conditions: (a) blurring levels with sigma varying from 0 to 20 and (b) noise levels with a mean of 5 and standard deviations varying from 0 to 20.}
    \label{fig:curve_metrics_vs_blur_and_noise}
\end{figure}

\begin{figure}[htbp]
    \centering
    \subfloat[]{\includegraphics[width=0.4\textwidth]{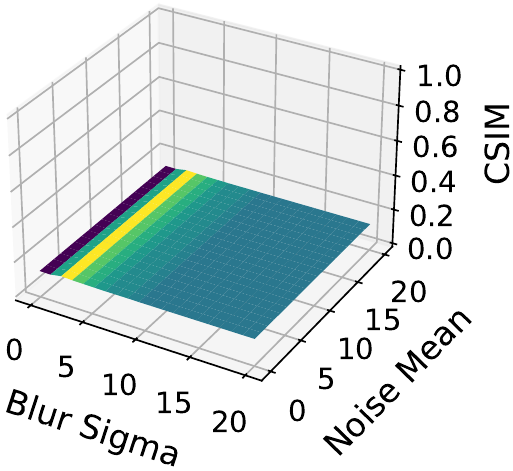}%
    \label{fig:3d_csim}}
    \hfill
    \subfloat[]{\includegraphics[width=0.4\textwidth]{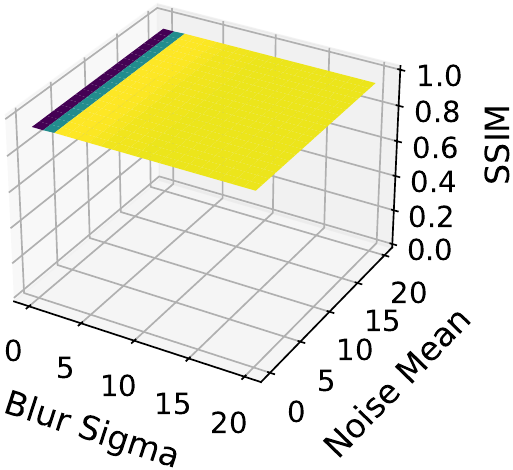}%
    \label{fig:3d_ssim}}
    \vfill
    \subfloat[]{\includegraphics[width=0.4\textwidth]{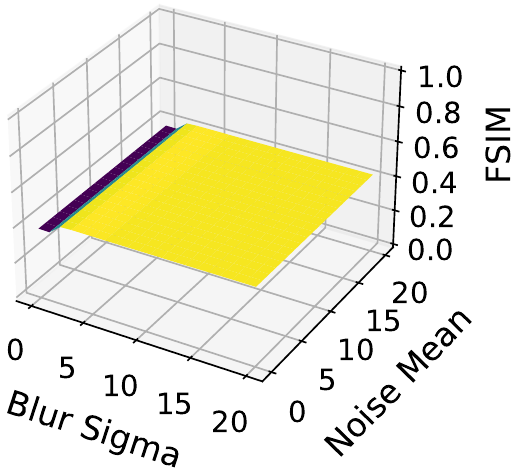}%
    \label{fig:3d_fsim}}
    \hfill
    \subfloat[]{\includegraphics[width=0.4\textwidth]{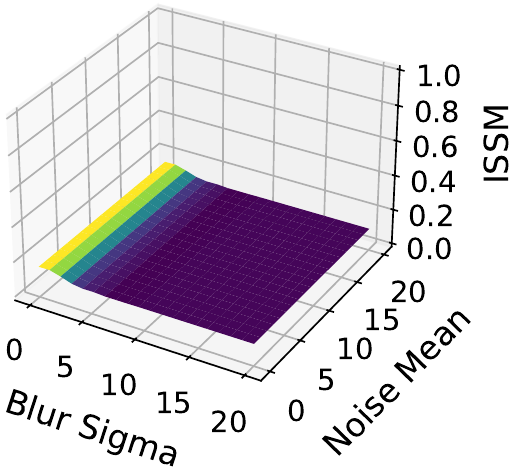}%
    \label{fig:3d_issm}}
    \caption{3D visualizations of image quality metrics as functions of blur sigma and noise mean: (a) CSIM, (b) SSIM, (c) FSIM and (d) ISSM.}
    \label{fig:3d_metrics}
\end{figure}

The results of these evaluations are presented in Figs. \ref{fig:curve_metrics_vs_blur_and_noise} and \ref{fig:3d_metrics}. Fig. \ref{fig:curve_metrics_vs_blur_and_noise} includes: \ref{fig:metrics_vs_blur} comparison between the metrics as functions of blur sigma, while \ref{fig:metrics_vs_noise} compares the metrics as functions of noise sigma. It can be observed that as the blur level increases, the similarity scores for all metrics decrease, indicating a degradation in image quality. This effect is particularly pronounced for CSIM and ISSM, which are sensitive to both blurring and noising. Both of the later performances are comparable as they demonstrate stable performances across different blur and noising levels. The stability of CSIM suggests its robustness to local variation in image. 

Further analysis is conducted through 3D visualizations of the metrics as functions of both blur and noise parameters, presented in Fig. \ref{fig:3d_metrics}. It reveal how the combined effects of noise and blur interact to influence the metric scores. For CSIM and ISSM, the surface plot indicates that the metric is very sensitive to noise and to blur, with higher values of noise significantly reducing the similarity score regardless of the blur level. FSIM shows a similar trend but with less pronounced effects and finally SSIM is hardly locating the difference between image under those noising and blurring conditions. 

\subsection{Video-based dynamic assessment}

\begin{figure*}[!t]
     \centering
     \subfloat[]{\includegraphics[width=\textwidth]{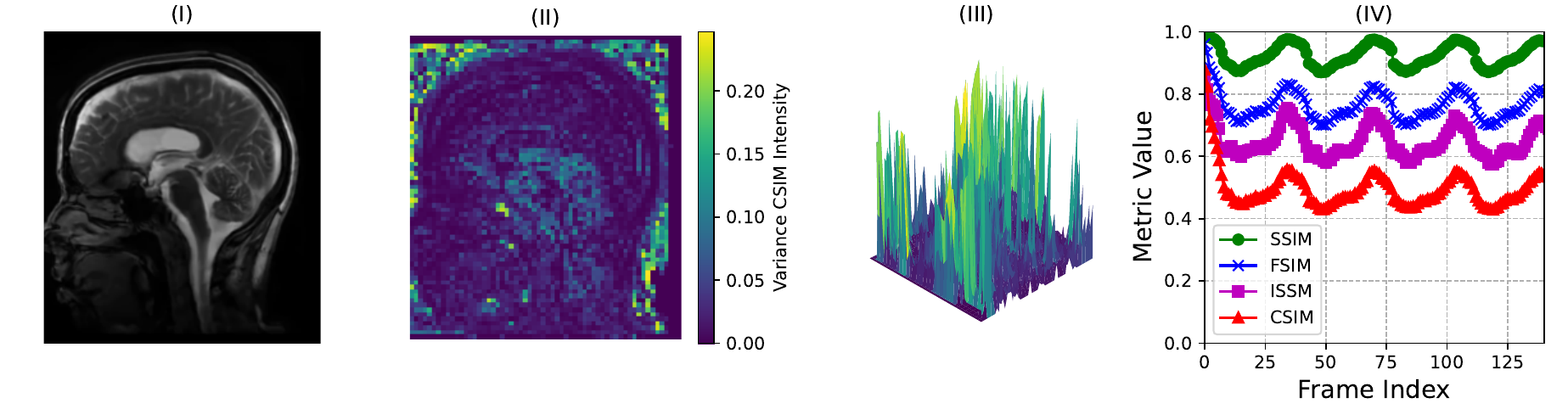}
     \label{fig:dynamic_analysis_medical}}
     \hfil
     \subfloat[]{\includegraphics[width=\textwidth]{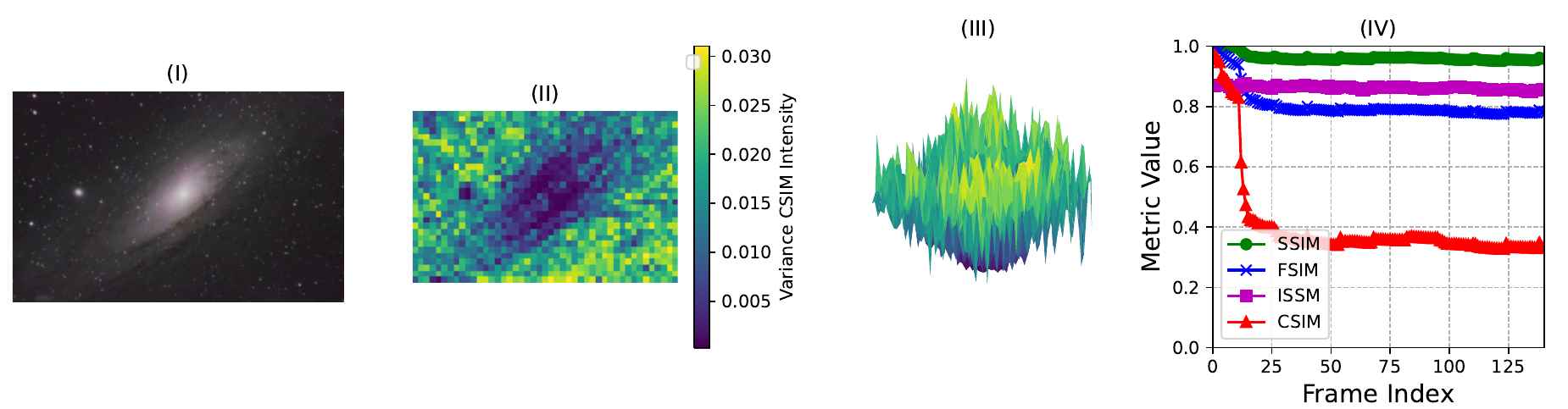}%
     \label{fig:dynamic_analysis_galaxy}}
     \caption{Dynamic evaluation of similarity metrics applied in two fields: Healthcare through MRI brain video and Astronomy through Andromeda Galaxy video: First frame fixed a reference then compared to subsequent frames: (I) reference frame, (II) Overall recorded changes in patches across all frames, (III) 3D visual of the highest similarity changes, (IV) Comparison between the metrics. CSIM is shown to be the most sensitive to localized changed fluctuating between 0.4 and 0.6 for the MRI video and 0.38 for the captured Andromeda galaxy, while SSIM gives consistently high similarity scores above 0.95, indicating its insensitivity to small changes.}
     \label{fig:dynamic_analysis_medical_astronomy}
 \end{figure*}

A dynamic assessment was conducted on a videos with minor changes and large static background (Fig. \ref{fig:dynamic_analysis_medical} for the MRI video scan of a human brain and Fig. \ref{fig:dynamic_analysis_galaxy} for the Andromeda Galaxy video, both videos can be found on CopulaSimilarity GitHub repository), with the primary objective of analyzing the effectiveness of the proposed CSIM metric in detecting subtle changes over time. The videos, which predominantly exhibits a small proportion of movement, were processed using the four similarity metrics. The initial frame of the video was selected as a reference and subsequent frames were compared against this reference frame to compute the similarity scores using each metric. The video frames were first resized to reduce from 4k resolution down to 1080p, enhancing processing efficiency and mostly due to FSIM high complexity and processing time and then converted to RGB format for uniformity across different metrics.

The similarity scores were calculated dynamically over frames, providing a visual representation of each metric's responds to changes in the videos. It was observed that the SSIM consistently yielded very high similarity scores, often exceeding 0.95, indicating its insensitivity to minor changes and noise in the MRI video. In contrast, the CSIM demonstrated a higher sensitivity to localized changed, with similarity scores mostly fluctuating significantly between 0.4 and 0.6 for the MRI video and 0.38 for the Andromeda Galaxy. Meanwhile, the FSIM and ISSM produced relatively similar results, with FSIM being slightly less sensitive compared to ISSM. Similarly for the astronomy application, CSIM kept delivering low similarity scores compared to the rest of the metrics (around 0.4).
 
This dynamic analysis highlights the distinct characteristics of each similarity metric when applied to medical imaging, particularly in video sequences where subtle motion or changes need to be detected. The sensitivity of CSIM to noise and its ability to capture finer details make it a valuable tool in medical diagnostics, especially in scenarios where the accurate detection of small changes is needed.

\subsection{CSIQ Dataset: performance evaluation}

\begin{figure}[ht]
    \centering    \includegraphics[width=0.8\textwidth]{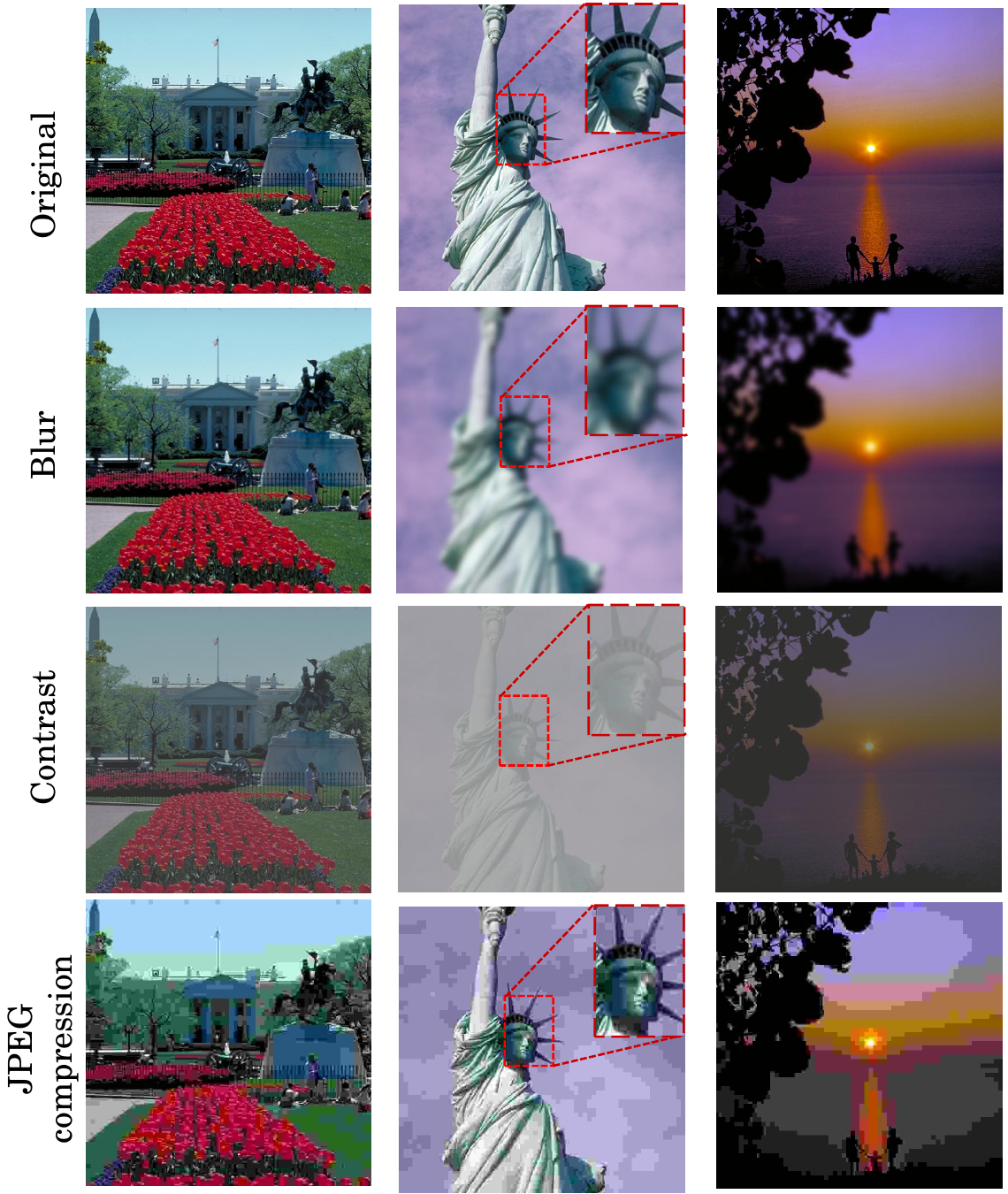}
    \caption{Example of three original images with their equivalent artificially distorted versions from CSIQ database using contrast, blur and two JPEG compression types with a magnification on the head of lady liberty for each distortion.}
    \label{fig:dataset_sample_visualization}
\end{figure}

The evaluation of the effectiveness of the proposed metric is conducted on a large dataset \citep{10.1117/1.3267105} while comparing its performance to the other metrics. Categorical Subjective Image Quality (CSIQ) database consists of originally taken pictures provided along with distorted versions based on six different types: JPEG  and JPEG-2000 compression, global contrast decrements, frequency noise, additive white Gaussian noise and Gaussian blurring (Fig. \ref{fig:dataset_sample_visualization}). For each distortion type, five levels of image alteration have been applied leading to a dataset that consists of 30 original images and 900 distorted versions (30 distortions per original image).

\begin{figure}[ht]
    \centering
    \includegraphics[width=0.7\textwidth]{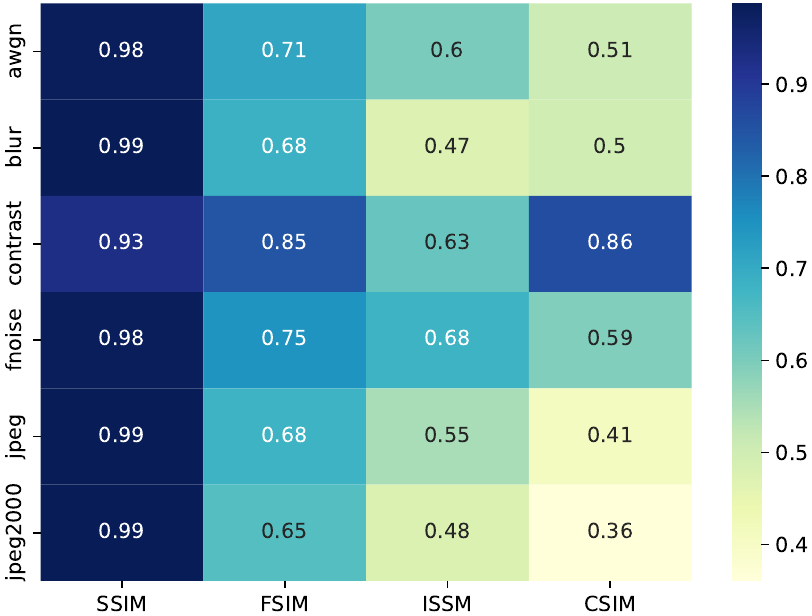}
    \caption{Metric performance evaluation with respect to each distortion type.}
    \label{fig:heatmap_average_scores}
\end{figure}

The heatmap presented in Fig.\ref{fig:heatmap_average_scores} provides a comparative view of the average similarity scores for various metrics—SSIM, FSIM, ISSM and CSIM—across different types of artificial image distortions, including Additive White Gaussian Noise (AWGN), Blur, Contrast Enhancement, Frequency Noise (FNoise), JPEG and JPEG2000 Compression.

SSIM kept delivering high scores when comparing the original image to every type of distortion. For AWGN and blur, FSIM, ISSM and CSIM show low scores indicating their robustness in locating small changes (CSIM being the lowest amongst the metrics). CSIM's performance is also comparatively lower in scenarios involving fnoise (0.59). Additionally, CSIM’s lower similarity scores in JPEG and JPEG2000 Compressions (0.36 and 0.41 respectively) underscore its heightened sensitivity to compression artifacts, making it particularly adept at identifying these distortions especially compared to FSIM returning relatively higher scores.  

\begin{figure*}[!t]
    \centering
    \includegraphics[width=\textwidth]{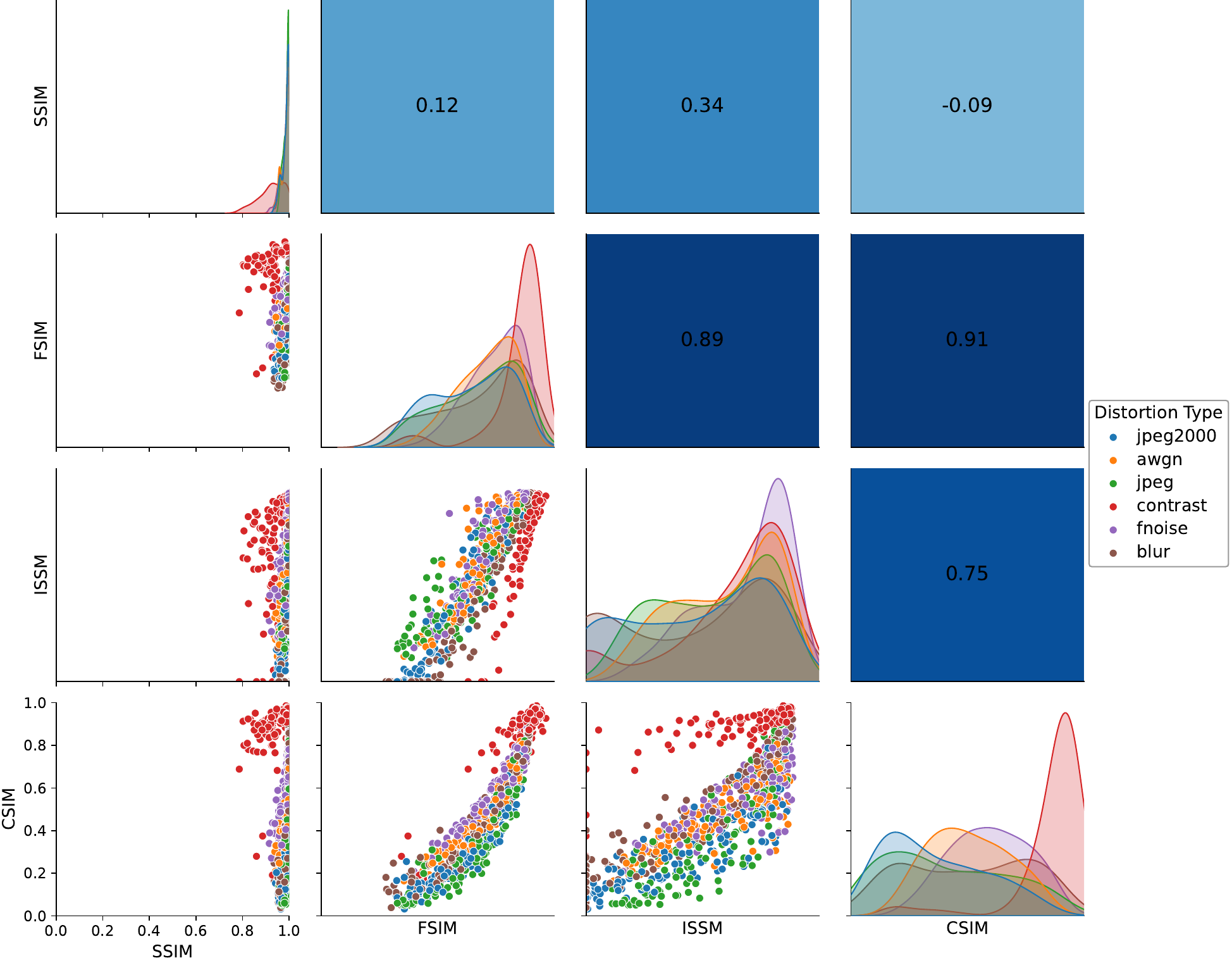}
    \caption{Pair-wise metric performance diagram with respect to each distortion type. Each subfigure compares two of the similarity metrics, with the scatter plots showing how scores from one metric relate to those from another. The diagonal subplots display histograms of each metric's scores, providing insight into their individual distributions. The rest of scatter plots in the lower triangle represent pairwise comparisons between the metrics, each figure contains a total of 900 points. Each point is corresponding to the score obtained between original and distorted version. The color coding indicates the type of distortion applied (\textcolor{blue}{blue}: jpeg2000, \textcolor{brown}{brown}: blur, \textcolor{orange}{orange}: awgn, \textcolor{green}{green}: jpeg, \textcolor{red}{red}: contrast, \textcolor{purple}{purple}: fnoise). The upper triangle represent the correlation score with blue color map ranging from light blue -0.09 (lowest correlation value) to dark blue 0.91 (highest correlation value).}
    \label{fig:similarity_scores_by_distortion_type}
\end{figure*}

Fig. \ref{fig:similarity_scores_by_distortion_type} presents a pairwise comparison of similarity scores obtained from four different image quality metrics. Each metric is assessed based on its ability to evaluate various types of image distortions. The comparison highlights both the relationship between these metrics and their sensitivity depending on the type of distortion applied to the images. The diagonal display the distribution of similarity scores for each metric individually in the form of kernel density estimates (KDE). These shows each metric's score variations across the different distorted images. SSIM, FSIM and ISSM displays score distributions in mid-to-high similarity values, which suggests that these metrics generally delivered high similarity, regardless of distortion type. The CSIM however, demonstrates an overall low score distribution, apart from contrast, reflecting its heightened sensitivity to image distortions.

The left-diagonal scatter plots represent pairwise comparisons of the metrics, with each point corresponding to the score delivered by an image comparison between the original and a distorted version. The color coding indicates the type of distortion applied. The provided scatter plots demonstrate the relationship between metrics. For instance, FSIM and ISSM show a relatively linear relationship, indicating that they tend to rate image similarly. Conversely, CSIM show a different pattern since it often assigns lower similarity scores compared to the other metrics. This suggests that CSIM is more sensitive to certain distortions, particularly when assessing subtle differences in image quality that SSIM, FSIM and ISSM might overlook. The correlation coefficients, displayed as color-coded boxes in the upper triangle, highlight how closely each pair of metrics correlates with one another where the lowest recorded correlation is between SSIM and CSIM (-0.09) and the highest is between CSIM and FSIM (0.91).

\section{Discussion}\label{sec:result_discussion}

\subsection{Algorithm complexity analysis}

A pseudocode representation of the algorithm is provided followed by a detailed analysis of its computational complexity while evaluating the computational efficiency of the proposed CSIM. The algorithm can be summarized in the pseudocode \ref{code:CSIM}. It has been executed on an Intel i9 11Gen Linux machine with 3 GPU's RTX A6000 48 Gb memory each and 128 GB of RAM.

\begin{algorithm}\label{code:CSIM}
\caption{CSIM pseudo-code}
    \begin{algorithmic}[1]
        \renewcommand{\algorithmicrequire}{\textbf{Input:}}
        \renewcommand{\algorithmicensure}{\textbf{Output:}}
        \REQUIRE image1, image2, patch\_size
        \ENSURE similarity matrix
        \\ \textit{Initialisation} :
        \STATE Initialize patch\_size
        \\ \textit{Feature Extraction and Copula Computation}
        \STATE patches1 $\leftarrow$ ExtractLocalFeatures(image1, patch\_size)
        \STATE patches2 $\leftarrow$ ExtractLocalFeatures(image2, patch\_size)
        \STATE local\_similarities $\leftarrow$ []
        \\ \textit{LOOP Process}
        \FOR {each (patch1, patch2) in (patches1, patches2)}
        \STATE features1 $\leftarrow$ Reshape(patch1)
        \STATE features2 $\leftarrow$ Reshape(patch2)
        \STATE copula1 $\leftarrow$ ComputeCopula(features1)
        \STATE copula2 $\leftarrow$ ComputeCopula(features2)
        \STATE euc\_distance $\leftarrow$ EuclideanDistance(copula1, copula2)
        \STATE similarity $\leftarrow$ Max(0, 1 - euc\_distance)
        \STATE Append similarity to local\_similarities
        \ENDFOR
        \\ \textit{Return Reshaped Similarity Matrix}
        \RETURN Reshape(local\_similarities)
    \end{algorithmic}
\end{algorithm}

The computational complexity of the CSIM algorithm can be presented as follows:

\begin{itemize}
    \item \textbf{ExtractLocalFeatures:} 
    this function extracts non-overlapping patches from the input image. Given an image of size \(M \times N\) and patch size \(P \times P\), the number of patches extracted is \(\frac{M}{P} \times \frac{N}{P}\). The complexity of extracting all patches is \(O\left(\frac{MN}{P^2}\right)\).

    \item \textbf{Reshape:}
    the reshape operation on a patch of size \(P \times P \times C\) (where \(C\) is the number of channels, typically 3 for RGB) has a complexity of \(O(1)\) as it does not involve data movement but only changes the view of the array.

    \item \textbf{ComputeCopula:}
    this function involves computing ranks and then the PPF:
    \begin{itemize}
        \item \textit{ComputeRanks:} sorting the pixel intensities for each patch has a complexity of \(O(P^2 \log P^2)\), which simplifies to \(O(P^2 \log P)\).
        
        \item \textit{PPF Normal:} applying the Percent Point Function (PPF) to convert ranks into copula values has a complexity of \(O(P^2)\).
    \end{itemize}

    Therefore, the overall complexity of computing the copula for a single patch is \(O(P^2 \log P)\).

    \item \textbf{EuclideanDistance:}
    Calculating the Euclidean distance between two vectors (each of size \(P^2 \times C\)) has a complexity of \(O(P^2)\).

    \item \textbf{Loop Over Patches:}
    The algorithm iterates over all patches and for each pair of corresponding patches from the two images, computes the copula and similarity. Given the number of patches \(\frac{MN}{P^2}\), the overall loop complexity is expressed in Eq. \ref{eq:loop_complexity}:
    \begin{equation}\label{eq:loop_complexity}
        O\left(\frac{MN}{P^2} \times (P^2 \log P + P^2)\right) = O(MN \log P)
    \end{equation}

\end{itemize}

\begin{figure}[!t]
    \centering
    \includegraphics[width=0.8\textwidth]{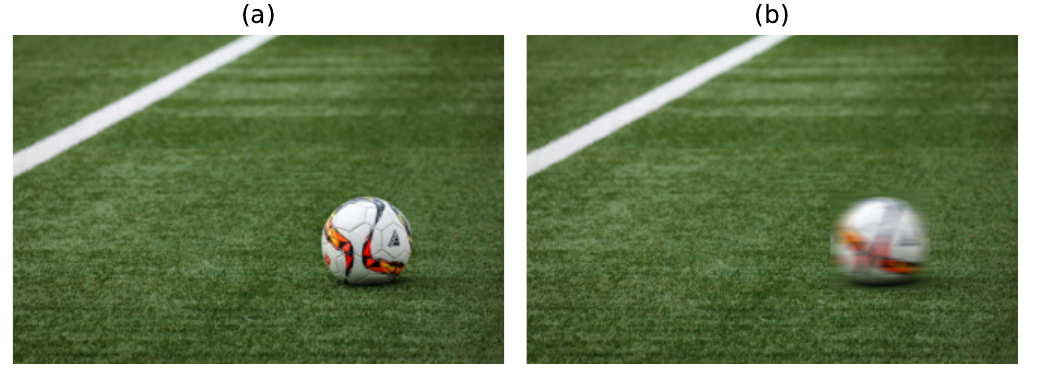}
    \caption{Very high resolution image selected for algorithm complexity analysis: (a) original image, (b) same image with distorted ball area}
    \label{fig:4k_resolution_img}
\end{figure}

Combining all the steps, the dominant term is \(O(MN \log P)\) which constitutes the overall time complexity of the algorithm. To evalutate Time complexity of the proposed method, an image of 4K resolution $(3888 \times 2592)$ has been selected. Afterward, a shutter blurring distortion has been applied on a specified rectangle to create localized change (Fig. \ref{fig:4k_resolution_img}).

\begin{figure}[!t]
    \centering
    \includegraphics[width=0.7\textwidth]{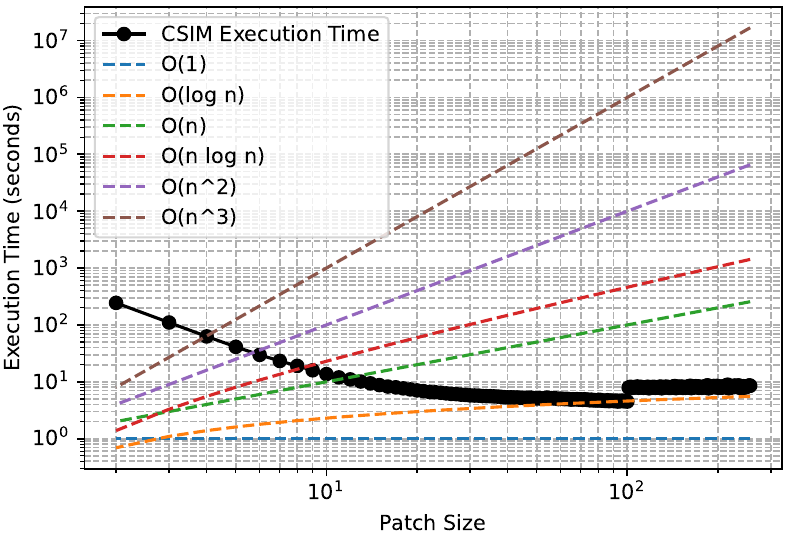}
    \caption{Complexity analysis of the CSIM similarity in comparison to other big O complexities. The less the patch size, the more execution time tends toward $O(n^3)$. Conversely, starting from patch size of 10, the complexity decreases to $O(n)$ until $O(\log n)$.}
    \label{fig:complexity_analysis}
\end{figure}

CSIM has been applied while looping over patch sizes increasing from 1 up to 256. With a patch size set to $1$ on the 4k resolution image, the algorithm took around 16 minutes to run and it hasn't been able to accurately spot the changes within the distorted area. However, with a patch size set to 256, the CSIM value decreases rapidly while taking 8 seconds to compute similarity map. Fig. \ref{fig:complexity_analysis} shows a visual of time requirement needed to run the algorithm on the high resolution image. Thus, the algorithm is primarily affected by the resolution size of the image and the logarithmic factor of the patch size. Nonetheless, decreasing patch size to very small windows and considering it on very high resolution images also lead to inaccuracies because pairs of patches at the pixel-level look very similar. One main limitation of the proposed approach lies in the choice of the right patch size with respect to the image resolution. Here, a default patch size of 8 has been set for all the images which allowed us to obtain the above results. According to the complexity curve (Fig. \ref{fig:complexity_analysis}), the patch size required for 4k resolution image need to be within a range of $10$ and $10^2$ to have a good balance between processing time and accuracy.

\section{Conclusion}
In this paper, one similarity metric for image assessment is proposed. The metric, namely CSIM, based on Gaussian Copula, is applied on small image patches to use the statistical dependencies between pixel intensities to accurately compare two images. CSIM computation can be grouped into four main actions: (1) patching: consists of breaking the image into small square patches and sorting the pixel for each color channel. (2) Copula patch: includes the computation of pixel ranks and CDF followed by PPF functions. (3) Copula vectors: groups the obtained copula patches into a linear vector specific for each image. (4) Similarity map: consist of Euclidean distance calculation between the two copula vectors. The proposed CSIM has proven its efficiency compared to the other metrics SSIM, FSIM and ISSM with a more nuanced capturing of localized changes between images. With regards to the conducted static experiments, where distortions have been applied on images (experiments under Gaussian blur and noise), it is observable that CSIM outperforms SSIM and FSIM, while also being comparable to ISSM (in some cases, ISSM is slightly better at capturing blur and CSIM slightly better at detecting Gaussian noise). To further analyse the efficiency of the proposed method, a dynamic testing has been performed, where high resolution videos have been collected with a small moving areas. The CSIM demonstrated high sensitivity and has been able to identify those local changes quite efficiently in comparison to the other metrics. Additionally, a CSIQ database composed of several original high resolution images and artificially distorted ones, has been selected to compare the sensitivity of the metrics. CSIM has show robustness in detecting differences between original and compressed images as well as detecting frequency noise, while also being comparable to ISSM in detecting blur and additive white Gaussian noise. To make the work accessible to the scientific community, a python package have been released for an easy use of CSIM under the Github repository: \url{https://github.com/safouaneelg/copulasimilarity}. It also contain all the steps and material to reproduce the results presented within this paper.


\bibliographystyle{unsrt}
\bibliography{references}



\end{document}